\documentclass{optica-article}
\usepackage{siunitx}
\usepackage{subcaption}
\journal{opticajournal} 

\articletype{Research Article}

\usepackage{lineno}

\begin{document}

\title{BeST: Beta splatted tomography for direct reconstruction from coherent imaging measurements}

\author{Johannes Gruen\authormark{1,*}, Philipp Paetzold\authormark{1}, Andreas Schropp\authormark{1}, Aknur Karabay\authormark{2,3}, Shahin Sepanlou\authormark{1}, Sebastian Eberle\authormark{9}, Martin Seyrich\authormark{1}, Felix Wittwer\authormark{1,5,6}, Imke Greving\authormark{7}, Silja Flenner\authormark{7}, Ken Vidar Falch\authormark{4}, Jan Garrevoet\authormark{4}, Manuel Guizar-Sicairos\authormark{2,3}, Christian G.
Schroer\authormark{1,8,9}, Johannes Hagemann\authormark{1}}

\address{
\authormark{1}Center for X-Ray and Nano Science (CXNS), Deutsches Elektronen-Synchrotron (DESY), Hamburg, Germany\\
\authormark{2}Paul Scherrer Institute (PSI), Villigen, Switzerland\\
\authormark{3}École Polytechnique Fédérale de Lausanne (EPFL), Lausanne, Switzerland\\
\authormark{4}Deutsches Elektronen-Synchrotron (DESY), Hamburg, Germany\\
\authormark{5}Physics Department, University of Siegen, Siegen, Germany\\
\authormark{6}currently at Karlsruhe Institute of Technology (KIT), Karlsruhe, Germany\\
\authormark{7}Institute of Materials Physics, Helmholtz-Zentrum Hereon, Hamburg, Germany\\
\authormark{8}Physics Department, Universität Hamburg, Hamburg, Germany\\
\authormark{9}Helmholtz Imaging, Deutsches Elektronen-Synchrotron (DESY), Hamburg, Germany\\
}

\email{\authormark{*}johannes.gruen@desy.de} 


\begin{abstract*}
	Conventional coherent X-ray tomography requires the phase of each projection to be reconstructed with high fidelity.
	However, individual reconstructed projections might be flawed, e.g. due to positioning errors, yielding inconsistent tomographic reconstructions.
	By directly coupling phase retrieval and tomography, consistency is enforced through a common object volume.
	But reconstructing a full volume demands substantial computational resources.
	 To reduce this demand, we introduce Beta Splatted Tomography (BeST), which represents the common object volume using compactly supported beta primitives with closed-form analytical projections, enabling efficient reconstruction directly from coherent scattering images.
	This representation substantially reduces the number of optimizable parameters compared to a voxel grid.
	We demonstrate BeST on experimental datasets from hard X-ray near-field holography and far-field ptychography measurements.
	In both cases, the joint technique improves reconstruction quality compared to two-step methods.
\end{abstract*}

\section{Introduction}
Understanding the internal structure of samples is essential to solving questions in biology  and  material science.
Imaging techniques, particularly X-ray tomography, reveal this structure, providing insights into pathological processes~\cite{Eckermann2020} and informing the optimization of functional materials, including catalysts~\cite{Meirer5055}.
Multimodal imaging enriches this structural information with complementary measurements, such as elemental distributions~\cite{Grun26}.
Hard x rays from synchrotron radiation sources are particularly suited to these investigations because their coherent and spectral properties enable access to the interior of specimens at the nanoscale.

Many specimens investigated at nanometer resolution absorb hard x rays only weakly and therefore provide little attenuation contrast.
The phase shifts they induce, however, provide a substantially stronger contrast mechanism.
Coherent imaging techniques exploit this effect by encoding phase variations into measurable intensity patterns.
Because the detector records only intensity, the phase information must be recovered computationally through a process known as phase retrieval~\cite{cloetensHolotomography1999,Fienup1978,Rodenburg2004}.

X-ray tomography based on coherent imaging methods acquires projections over a range of rotation angles, which are combined to reconstruct the 3D object.
In this work, we consider near-field holotomography and far-field ptychotomography.
Near-field holotomography is a full-field method that combines phase retrieval with tomography to recover 3D information~\cite{cloetensHolotomography1999}.
X-ray ptychography is a scanning far-field method that exploits overlapping illuminated regions for phase retrieval~\cite{Rodenburg2004,thibaultHighResolutionScanningXray2008}.
Combined with tomography it enables three-dimensional nanoscale imaging~\cite{dierolfPtychographicXrayTomography2010}, including 3D reconstructions of integrated circuits~\cite{hollerIntegratedCircuits2017,Aidukas2024}.
These two methods typically use a similar reconstruction scheme: projections are reconstructed independently and subsequently combined into a volume~\cite{guizarPhaseTomography2011}.
This approach does not enforce consistency with a 3D object during phase retrieval, allowing discrepancies between projections to propagate into the tomographic reconstruction.
Joint reconstruction instead couples the measurements through a common object representation, enforcing tomographic consistency and providing additional constraints that can stabilize phase retrieval under measurement uncertainties.

Previous work has formulated ptychographic phase retrieval and tomography as a joint inverse problem, using direct gradient-based optimization or alternating schemes that couple ptychographic and tomographic subproblems~\cite{gursoyDirectCouplingTomography2017,ramosDirect3DTomographic2019,aslanJointPtychoTomography2019,changIterativeJointPtychography2019,barutcuSimultaneous3DXray2020}.
Kahnt~\textit{et al.} demonstrated that coupling ptychographic and tomographic reconstruction improves spatial resolution in experimental ptychotomographic data~\cite{kahntCoupledPtychographyTomography2019}.
In near-field holography, Ruhlandt and colleagues coupled phase retrieval and algebraic tomography through iterative reprojection~\cite{ruhlandtThreeDimensionalPhaseRetrieval2014} and subsequently introduced a 3D propagation scheme under the weak-object approximation~\cite{ruhlandtThreeDimensionalPropagation2016}.
Automatic-differentiation frameworks have enabled optimization of 3D object models from near- and far-field intensity measurements, including multislice propagation to account for wave propagation within the specimen when the projection approximation is insufficient~\cite{duThreeDimensionsTwo2020,duAdorymMultiplatformGeneric2021}.

A major computational challenge for joint reconstruction is the number of optimizable parameters when the object is represented on a dense voxel grid.
For the ptychographic application considered here, a grid covering the reconstructed field of view (FOV) at the chosen sampling would contain approximately $4800 \times 4800 \times 700$ voxels.
Independently optimizing both components of the complex refractive index in a voxel representation would introduce more than $3.2 \times 10^{10}$ scalar unknowns.
Storing these arrays, their gradients, and optimizer states, together with repeatedly evaluating the forward model, places substantial demands on memory and computation.

Continuous representations offer alternatives to optimizing an independent value at every voxel.
Zha~\textit{et al.} represent the attenuation coefficient as a continuous function of spatial coordinates parameterized by a neural network, whose parameters are optimized against measured projections~\cite{zhaNAFNeuralAttenuation2022b}.
An alternative uses Gaussian primitives whose analytical projections permit efficient forward projection~\cite{zhaR2GaussianRectifyingRadiative2024}.
Recently, implicit neural representations (INR) have also enabled joint reconstruction directly from inline holographic measurements~\cite{gruenXRayNearFieldHolotomography2026a}.
INRs provide a compact object parameterization, but evaluating their projections can still require many network queries along integration paths.
Combining a compact representation with analytical projection is therefore an attractive route to reducing the computational cost of joint phase retrieval and tomography.

Here, we introduce Beta Splatted Tomography (BeST), a method for joint phase retrieval and tomography using a continuous complex refractive-index representation built from compactly supported beta primitives.
Following Liu~\textit{et al.}~\cite{liuDeformableBetaSplatting2025}, we adopt a beta kernel with an optimizable shape parameter that adjusts its radial profile in addition to the spatial extent and orientation controlled by the primitive scales and rotation.
Building on their inverse Abel construction, we derive an explicit closed-form volumetric kernel to represent and evaluate the 3D refractive-index fields.
We incorporate analytical line-integral projections of these primitives into differentiable forward models for near-field inline holography and far-field ptychography.
The projected refractive-index contrast determines the complex transmissivity, which is propagated to predict the measured intensities and optimize the 3D object.

We evaluate BeST on experimental hard X-ray datasets acquired at PETRA~III at DESY (Hamburg, Germany): a foraminifera specimen measured by near-field holotomography at the nano-imaging endstation of beamline P05~\cite{flennerHardXrayNanoholotomography2020b}, operated by Helmholtz-Zentrum Hereon, and a commercial microchip measured by limited-angle ptychotomography at the PtyNAMi instrument at beamline P06~\cite{schroppPtyNAMiPtychographicNanoanalytical2020}.

For the ptychographic dataset, residual scan-position errors after position correction prevent a usable tomographic reconstruction with the evaluated sequential workflow, whereas BeST recovers the microchip structure with an object-representation compression factor of 15 relative to a dense voxel representation.
For near-field holography, reconstruction requires only 30~minutes on a single GPU, compared to approximately 24~hours for the INR approach, and achieves a compression factor of 270 relative to a dense voxel representation.
The achievable compression factor depends on the specimen.

\section{Coherent Imaging Forward Models}
\begin{figure}
	\includegraphics[width=\linewidth]{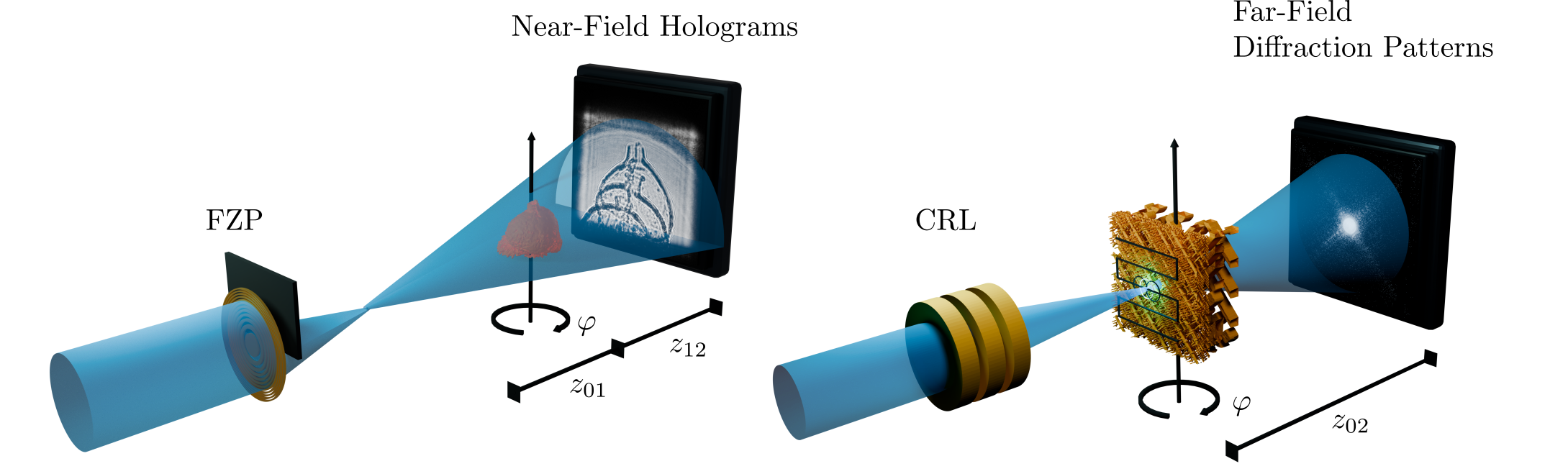}
	\caption{Schematic representation of the two coherent X-ray imaging configurations.
		Left: Inline holography using a Fresnel zone plate (FZP) to focus the incident beam.
		The sample is positioned in the diverging cone beam at a distance $z_{01}$ downstream of the focus.
		After interacting with the sample, the x rays propagate over the sample-to-detector distance $z_{12}$ before their intensity is recorded.
		Right: Far-field ptychography using compound refractive lenses (CRL) to focus the incident beam.
		The sample is positioned slightly downstream of the focus and scanned transversely through the illumination, producing overlapping illuminated regions.
		The far-field detector records one diffraction pattern at each scan position.
		In both configurations, measurements are acquired at multiple sample rotation angles $\varphi$ to obtain holotomographic and ptychotomographic datasets, respectively.}
	\label{fig:experimental-setup}
\end{figure}

In coherent imaging, the complex refractive index of the specimen is encoded in the phase and amplitude of the transmitted wavefield. 
It is given by~\cite{paganinCoherentXrayOptics2006a}
\begin{align}
	n(\mathbf{r})
	=1-\delta(\mathbf{r})+\mathrm{i}\beta(\mathbf{r}),
	\qquad
	\delta(\mathbf{r})\geq 0,
	\quad
	\beta(\mathbf{r})\geq 0,
	\label{eq:refractive-index}
\end{align}
where $\delta$ is the refractive-index decrement, $\beta$ is the extinction coefficient, and $\mathbf{r}\in\mathbb{R}^3$ denotes a position in the object.
The forward models below relate the complex refractive index to the measured intensities.
Under the projection approximation, we use the line integrals of $\delta$ and $\beta$ at projection angle $\varphi$~\cite{paganinCoherentXrayOptics2006a}:
\begin{align}
	D_{\varphi}(\mathbf{x})
	&=\int_{\ell_{\varphi,\mathbf{x}}}\delta(\mathbf{r})\,\mathrm{d}s,
	&A_{\varphi}(\mathbf{x})
	&=\int_{\ell_{\varphi,\mathbf{x}}}\beta(\mathbf{r})\,\mathrm{d}s.
	\label{eq:refractive-line-integrals}
\end{align}
Here, $\ell_{\varphi,\mathbf{x}}$ is the ray through projection-plane coordinate $\mathbf{x}\in\mathbb{R}^2$.
The corresponding sample transmission function relates the projected refractive-index fields to the complex wavefield, as in refractive phase-retrieval formulations~\cite{wittwerRefractiveFramework2022}:
\begin{align}
	T_{\varphi}(\mathbf{x})
	&=\exp\left[
		\mathrm{i}k\int_{\ell_{\varphi,\mathbf{x}}}
		\left(n(\mathbf{r})-1\right)\,\mathrm{d}s
	\right]
	=\exp\left[
		-\mathrm{i}kD_{\varphi}(\mathbf{x})
		-kA_{\varphi}(\mathbf{x})
	\right],
	\label{eq:sample-transmission}
\end{align}
where $k=\frac{2\pi}{\lambda}$ is the wavenumber with wavelength $\lambda$.

	We first describe the near-field and far-field forward models, then introduce the beta-splat representation used to reconstruct the 3D object directly from the measured intensities.

\subsection{Inline Near-Field Holography}
Inline near-field holography is a full-field technique in which an extended probe illuminates the specimen (Fig.~\ref{fig:experimental-setup}, left).
Assuming that the probe $P$ is constant during a measurement, the exit wave is modeled as
\begin{align}
	\Psi_{\mathrm{exit},\varphi}(\mathbf{x})
	=T_{\varphi}(\mathbf{x}) P(\mathbf{x}).
	\label{eq:near-field-exit-wave}
\end{align}
Propagation is described by the Fresnel operator
\begin{align}
	\mathcal{D}_{\mathrm{Fr}_{\mathrm{eff}}}\left(\Psi\right)
	=
	\mathcal{F}^{-1} \left(
		\exp\left(-\mathrm{i}\pi \frac{\| \boldsymbol{\nu} \|_2^2}{\mathrm{Fr}_{\mathrm{eff}}}
		\right)
		\mathcal{F}\left(\Psi\right)
	\right),
	\label{eq:fresnel}
\end{align}
where $\mathcal{F}$ and $\mathcal{F}^{-1}$ denote the two-dimensional Fourier transform and its inverse, $\boldsymbol{\nu}$ is the spatial frequency, and $\mathrm{Fr}_{\mathrm{eff}}$ is the effective Fresnel number.
The detector records only the intensity of the propagated wavefield:
\begin{align}
	\mathcal{I}^{\mathrm{NF}}_{\varphi}(\mathbf{x})
	=
	\left|
	\mathcal{D}_{\mathrm{Fr}_{\mathrm{eff}}}\left(
		\Psi_{\mathrm{exit},\varphi}(\mathbf{x})
	\right)
	\right|^2.
	\label{eq:near-field-intensity}
\end{align}
The Fresnel scaling theorem~\cite{paganinCoherentXrayOptics2006a} maps the diverging illumination to an equivalent parallel-beam geometry; the coordinate transformation and effective Fresnel number are given in Appendix~\ref{app:fresnel-scaling}.
In near-field holography, only a single intensity image is acquired per projection angle, providing less measurement redundancy for jointly estimating additional unknowns. 
We therefore assume a constant probe, which can be accounted for by flat-field division, and do not optimize an additive detector background. 
In contrast, the overlapping diffraction measurements acquired in ptychography provide additional constraints that enable joint refinement of the probe and background, as described in the following section.
This can be done in ptychography, as explained in the next part.

	\subsection{Far-Field Ptychography}

	In ptychography, a localized probe is scanned across overlapping sample regions (Fig.~\ref{fig:experimental-setup}, right)~\cite{rodenburgHardXrayLensless2007,thibaultHighResolutionScanningXray2008}.
	At scan position $\mathbf{t}_{j,\varphi}$, the probe is
	\begin{align}
		P_{j,\varphi}(\mathbf{x})
		&=
		P_{\varphi}\left(\mathbf{x}-\mathbf{t}_{j,\varphi}\right),
		\label{eq:shifted-probe-modes}
	\end{align}
	i.e. the probe is assumed to remain unchanged as it is translated across the sample.	
	The corresponding exit wave is
	\begin{align}
		\Psi_{j,\varphi}^{\mathrm{exit}}(\mathbf{x})
		&=
		T_{\varphi}(\mathbf{x})P_{j,\varphi}(\mathbf{x}).
		\label{eq:multimode-exit-wave}
	\end{align}
	Under far-field propagation, the predicted diffraction intensity is therefore
	\begin{align}
		\mathcal{I}^{\mathrm{FF}}_{j,\varphi}(\boldsymbol{\nu})
		&=
		a_{j,\varphi}^2
		\left|
		\mathcal{F}\left[
		\Psi^{\mathrm{exit}}_{j,\varphi}
		\right](\boldsymbol{\nu})
		\right|^2
		+
		\mathcal{I}_{\mathrm{bg},\varphi}(\boldsymbol{\nu}),
		\label{eq:multimode-ptychographic-intensity}
	\end{align}
	Here, $\boldsymbol{\nu}$ denotes a discrete Fourier-grid coordinate corresponding to a detector pixel, and $\mathcal{F}$ denotes the discrete Fourier transform.
	The factor $a_{j,\varphi}$ accounts for variations in the incident illumination amplitude between scan positions, and $\mathcal{I}_{\mathrm{bg},\varphi}$ denotes the additive detector background, which accounts for air scattering and other incoherent scattering effects.
	A unique 3D object determines the transmissivity at every projection angle, while the probe, detector background, and scan-position corrections are parameterized separately for each angle.
	The implementation uses mutually incoherent probe modes to account for partial coherence and illumination instabilities, as detailed in Appendix~\ref{app:probe-modes}.

\section{Beta-Splat Object Representation and Analytical Projection}\label{sec:splatting}
\begin{figure}
	\includegraphics[width=\linewidth]{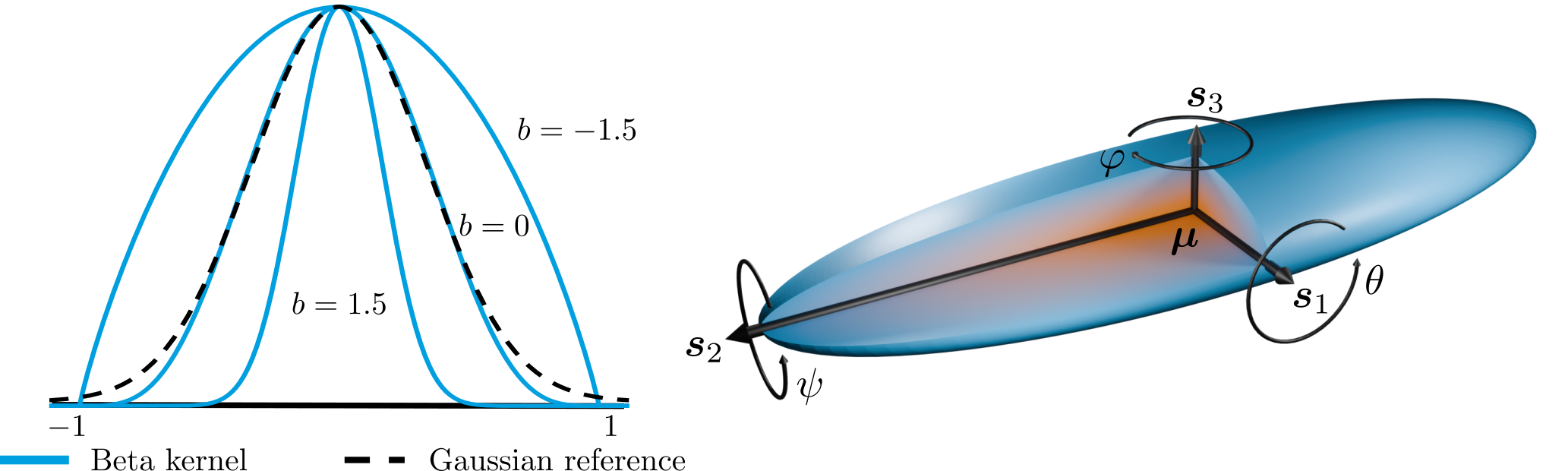}
	\caption{Left: One-dimensional cross-sections of the projected beta kernel for $b=-1.5$, $0$, and $1.5$ (solid blue), with a Gaussian reference (dashed orange).
	Increasing $b$ concentrates the profile toward its center.
	Right: Geometry of an anisotropic beta primitive, showing its center $\boldsymbol{\mu}$, semiaxis lengths $s_1$, $s_2$, and $s_3$, and orientation angles $\varphi$, $\theta$, and $\psi$.
	The color gradient illustrates the volumetric kernel profile $\mathcal{B}_3$, whose radial concentration is controlled by the shape parameter $b$.
	The orientation is parameterized by a unit quaternion $\mathbf{q}$ in the reconstruction.
	}
	\label{fig:beta}
\end{figure}
\begin{figure}
	\includegraphics[width=\linewidth]{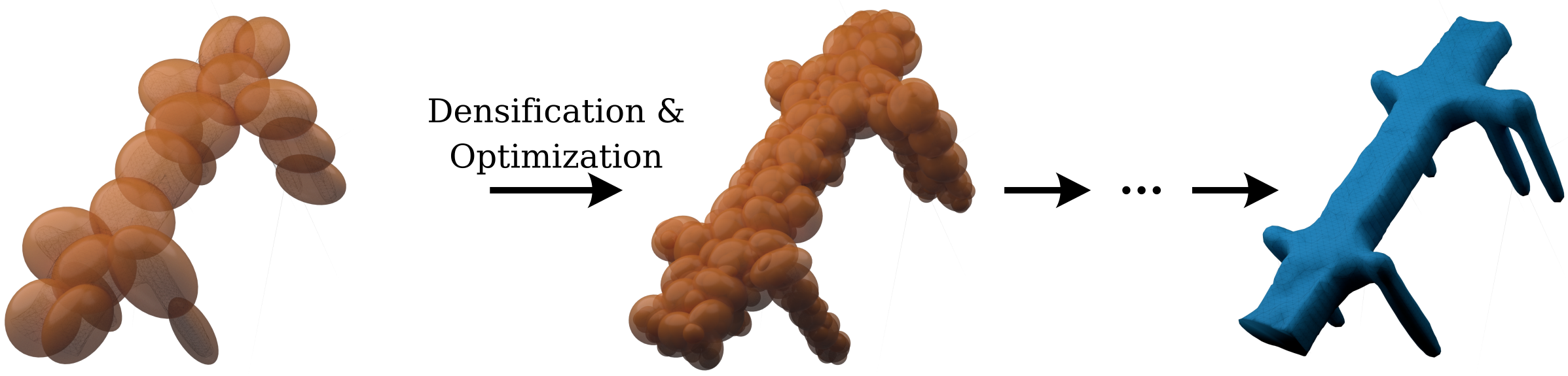}
	\caption{Schematic of primitive optimization in BeST.
	An initial set of primitives is optimized to obtain a coarse representation of the object (left).
	Subsequent optimization and splitting increase the number of primitives and yield a refined representation (center) that approximates the target object after more iterations (right).}
	\label{fig:optimization}
\end{figure}

We represent the 3D refractive-index distribution as a sum of anisotropic beta primitives (Fig.~\ref{fig:beta}).
Each primitive has an optimizable position, orientation, spatial extent, shape parameter, and two coefficients describing its refractive and absorptive contributions.
Splitting primitives during optimization increases the flexibility of the representation, as illustrated in Fig.~\ref{fig:optimization}.

Analytical integration of these primitives yields projections of the refractive-index fields.
These projections determine the specimen transmissivity at each viewing angle for either imaging model introduced in the previous section.

\subsection{Primitive Parameters and Object Fields}
A beta-splatting primitive $B_i$ (Fig.~\ref{fig:beta}) is parameterized by
\begin{align}
	B_i
	=
	\left(
		\boldsymbol{\mu}_i,
		\mathbf{q}_i,
		\mathbf{s}_i,
		b_i,
		\tilde{\delta}_i,
		\tilde{\beta}_i
	\right),
	\label{eq:beta-primitive}
\end{align}
where $\boldsymbol{\mu}_i\in\mathbb{R}^3$ is its center, $\mathbf{q}_i\in\mathbb{S}^3$ is a unit quaternion defining its orientation, $\mathbf{s}_i\in\mathbb{R}_{+}^3$ contains the semiaxis lengths of its ellipsoidal support, $b_i\in\mathbb{R}$ is its shape parameter, and $\tilde{\delta}_i,\tilde{\beta}_i\geq0$ are its refractive and extinction weights, respectively.
The ellipsoidal shape matrix is
$
	\boldsymbol{\Sigma}_i
	=
	\mathbf{R}\left(\mathbf{q}_i\right)
	\operatorname{diag}\left(\mathbf{s}_i^2\right)
	\mathbf{R}\left(\mathbf{q}_i\right)^{\mathsf T},
	$
where $\mathbf{R}(\mathbf{q}_i)\in\mathrm{SO}(3)$ is the rotation matrix associated with $\mathbf{q}_i$.
\subsection{Projected and Volumetric Beta Kernels}

We adopt the compactly supported beta kernel introduced by Liu~\textit{et al.}~\cite{liuDeformableBetaSplatting2025}.
We first prescribe the projected kernel, which is evaluated repeatedly during reconstruction, and then obtain its volumetric counterpart through the inverse Abel transform.
The inverse Abel transform allows the reconstruction of a 3D, rotationally symmetric distribution from its 2D projection, provided that the distribution satisfies suitable regularity and integrability conditions.
Let $r$ denote the normalized radial coordinate in the projection plane.
The projected two-dimensional kernel is defined as
\begin{align}
	\mathcal{B}_{2}\left(r;b\right)
	&=
	\begin{cases}
		\left(1-r^2\right)^{p(b)}, & 0 \leq r \leq 1,\\
		0, & r>1,
	\end{cases}
	& p(b)&=\max\left\{4\exp(b),p_{\min}\right\},
	\label{eq:beta-kernel-2d}
\end{align}
where the implementation uses $p_{\min}=0.55$.
Above this lower bound, increasing the shape parameter $b$ concentrates the projected profile toward its center at fixed support and unit peak, whereas decreasing $b$ broadens it (Fig.~\ref{fig:beta}).

The projected kernel describes the density accumulated along a ray, rather than the local density within the primitive.
For a radially symmetric volumetric kernel, this line integral is the Abel transform:
\begin{align}
    \mathcal{B}_2(r;b)
    =2\int_r^1
    \mathcal{B}_3(R;b)\frac{R}{\sqrt{R^2-r^2}}\,\mathrm{d}R.
    \label{eq:abel-projection}
\end{align}
The inverse Abel transform recovers the radial 3D density whose line integrals reproduce the prescribed 2D profile, ensuring consistency between volumetric evaluation and projection.
Here, we derive and use an explicit closed-form kernel expression to evaluate the 3D refractive-index fields and their volumetric regularization terms.
For unit support radius, the inverse transform is
\begin{align}
	\mathcal{B}_{3}\left(R;b\right)
	=
	-\frac{1}{\pi}
	\int_R^1
	\frac{\partial \mathcal{B}_{2}\left(r;b\right)}{\partial r}
	\frac{\mathrm{d}r}{\sqrt{r^2-R^2}},
	\qquad 0\leq R\leq1,
	\label{eq:inverse-abel}
\end{align}
where $R$ is the normalized 3D radial coordinate.
Evaluating Eq.~\eqref{eq:inverse-abel} gives the following closed form (see Appendix~\ref{app:abel-derivation} for the derivation):
\begin{align}
	\mathcal{B}_{3}\left(R;b\right)
	&=
	\begin{cases}
		\displaystyle
		\frac{\Gamma\left(p(b)+1\right)}
		{\sqrt{\pi}\,\Gamma\left(p(b)+\tfrac{1}{2}\right)}
		\left(1-R^2\right)^{p(b)-\frac{1}{2}},
		&0\leq R\leq1,\\[2mm]
		0, &R>1.
	\end{cases}
	\label{eq:beta-kernel-3d}
\end{align}
Here, $\Gamma$ denotes the gamma function, which extends the factorial to non-integer arguments and satisfies $\Gamma(m+1)=m!$ for non-negative integers $m$.
The floor $p_{\min}=0.55>\tfrac{1}{2}$ ensures that the volumetric kernel is non-negative, bounded, and continuous at its support boundary.
For each primitive, we map its ellipsoidal support to the unit ball and define
\begin{align}
	R_i(\mathbf{r})
	&=\sqrt{
		(\mathbf{r}-\boldsymbol{\mu}_i)^{\mathsf T}
		\boldsymbol{\Sigma}_i^{-1}
		(\mathbf{r}-\boldsymbol{\mu}_i)},
	& K_i(\mathbf{r})&=\mathcal{B}_3\left(R_i(\mathbf{r});b_i\right).
	\label{eq:volumetric-primitive}
\end{align}
Using the volumetric kernels $K_i$, the refractive-index decrement $\delta$ and extinction coefficient $\beta$ are represented by 
\begin{align}
	\delta(\mathbf{r})&=\sum_i\tilde{\delta}_i K_i(\mathbf{r}),
	&\beta(\mathbf{r})&=\sum_i\tilde{\beta}_i K_i(\mathbf{r}).
	\label{eq:continuous-object}
\end{align}
The coefficients $\tilde{\delta}_i$ and $\tilde{\beta}_i$ weight each primitive's contribution; their relation to peak density depends on the kernel normalization.
\subsection{Analytical Projection}
The 3D volume is related to the 2D projection as follows:
We evaluate the object line integrals using orthographic projection, with parallel rays at each rotation angle.
Let $\mathbf{W}_{\varphi}\in\mathbb{R}^{2\times3}$ be the orthographic projection matrix for angle $\varphi$, with orthonormal rows, and let $\mathbf{t}_{\varphi}\in\mathbb{R}^2$ denote the corresponding projection-plane offset.
Then the projected center becomes $\boldsymbol{\mu}'_{i,\varphi}=\mathbf{W}_{\varphi}\boldsymbol{\mu}_i+\mathbf{t}_{\varphi}$ and the projected shape matrix is $\boldsymbol{\Sigma}'_{i,\varphi}=\mathbf{W}_{\varphi}\boldsymbol{\Sigma}_i\mathbf{W}_{\varphi}^{\mathsf T}$.
For a projection-plane coordinate $\mathbf{x}\in\mathbb{R}^2$, the normalized radial distance to primitive $i$ is
$
	r_{i,\varphi}\left(\mathbf{x}\right)
	=
	\sqrt{
		\left(\mathbf{x}-\boldsymbol{\mu}'_{i,\varphi}\right)^{\mathsf T}
		\left(\boldsymbol{\Sigma}'_{i,\varphi}\right)^{-1}
		\left(\mathbf{x}-\boldsymbol{\mu}'_{i,\varphi}\right)
	}.
	$
Integration along the projection direction introduces the geometric factor
	$w_{i,\varphi}
	=
	\sqrt{
		\frac{\det\left(\boldsymbol{\Sigma}_i\right)}
		{\det\left(\boldsymbol{\Sigma}'_{i,\varphi}\right)}
	}.
	$
This factor accounts for the primitive's thickness along the viewing direction.
The line integrals of the refractive-index are therefore
\begin{align}
	D_{\varphi}(\mathbf{x})
	&=\sum_i \tilde{\delta}_i w_{i,\varphi}
	\mathcal{B}_2\left(r_{i,\varphi}(\mathbf{x});b_i\right),
	&
	A_{\varphi}(\mathbf{x})
	&=\sum_i \tilde{\beta}_i w_{i,\varphi}
	\mathcal{B}_2\left(r_{i,\varphi}(\mathbf{x});b_i\right).
	\label{eq:projected-object}
\end{align}
These expressions evaluate the object line integrals directly from the primitive's parameters, without sampling a dense volume along each ray, yielding a fast projection of the volume.
The resulting projections determine the complex transmissivity used by both imaging models introduced above.

\section{Experimental Methods}
We evaluate BeST on two complementary experimental datasets.
The near-field holography experiment assesses reconstruction quality and computational efficiency relative to an existing joint reconstruction method and a sequential workflow.
The far-field ptychography experiment examines reconstruction under geometric uncertainty by comparing joint optimization with independent two-step projection reconstruction followed by tomography.

\subsection{Reconstruction Procedure}
\label{sec:reconstruction-procedure}
The beta-splatting is implemented as a differentiable CUDA kernel adapted from the Gaussian-splatting of Zha et al.~\cite{zhaR2GaussianRectifyingRadiative2024}.
We replace the Gaussian primitive with the beta primitive and assign independent refractive and absorptive weights to each primitive.
Both imaging models are coupled to this projector, allowing gradients from the intensity-data loss to propagate to the parameters of the 3D object.

We optimize the beta-splat parameters using Adam~\cite{kingmaAdamMethodStochastic2015} by minimizing the squared $\ell_2$ distance between the square roots of the predicted and measured intensities~\cite{godardNoiseModelsLow2012}.

We additionally apply 3D total-variation (TV) regularization and $\ell_1$ regularization.
To estimate the TV term efficiently, we query the beta-splat representation on randomly selected subvolumes and calculate spatial finite differences within each subvolume.
The overall training objective is
\begin{align}
	\mathcal{L}
	&=
	\left\|
	\sqrt{\mathcal{I}}
	-
	\sqrt{\mathcal{I}_{\mathrm{det}}}
	\right\|_2^2
	+
	\lambda_{\mathrm{TV}}
	\mathcal{L}_{\mathrm{TV}}\left(V_{\mathrm{TV}}\right)
	+ \lambda_{\ell_1}
	\mathcal{L}_{\ell_1}\left(V_{\ell_1}\right),
	\label{eq:optimization-loss}
\end{align}
where $\mathcal{I}$ denotes the predicted holographic or ptychographic intensity and $\mathcal{I}_{\mathrm{det}}$ denotes the measured detector intensity.
The TV and $\ell_1$ penalties are evaluated on sampled subvolumes $V_{\mathrm{TV}}$ and $V_{\ell_1}$, with weights $\lambda_{\mathrm{TV}}$ and $\lambda_{\ell_1}$, respectively.

Figure~\ref{fig:optimization} illustrates how parameter refinement and densification improve the representation.
The shape parameter is initialized at $b=0$, giving $p(0)=4$ and the near-center profile $\mathcal{B}_2(r;0)\approx\exp(-4r^2)$.
Training begins with $10^5$ beta primitives placed randomly in space.
After a warm-up phase, primitives are split or cloned according to their position gradients and spatial extent, and weak primitives are pruned.
The detailed schedule and learning rates are given in Appendix~\ref{app:optimization-settings}.

\subsection{Near-Field Holotomography}
\label{sec:near-field-methods}
BeST was evaluated on a foraminifera specimen measured at the nanoimaging setup of beamline P05 at PETRA~III at DESY (Hamburg, Germany), operated by Helmholtz-Zentrum Hereon.
A setup based on a Fresnel zone plate was used as described by Flenner~\textit{et al.}~\cite{flennerHardXrayNanoholotomography2020b}; a schematic is shown in Fig.~\ref{fig:experimental-setup}.
The measurements were conducted at \qty{17}{\keV} across $1584$ angles equally distributed over approximately \ang{180} using a detector with a pixel size of \qty{6.5}{\micro\meter}.
The geometry corresponded to a Fresnel number of \num{9.979e-4} and a magnification of $29.68$.
We acquired 50 flat-field measurements without the specimen in the beam before and after measuring the projections.

Following the artifact-suppressing reconstruction method (ASRM) preprocessing procedure~\cite{doraArtifactsuppressingReconstructionStrongly2024}, we used principal component analysis of the flat-field measurements to construct synthetic flat fields and divided the measured holograms by the corresponding flat fields.
The corrected holograms were used as input to the reconstruction.

For near-field holography, we combine the projected object with a differentiable Fresnel propagator implemented in PyTorch, using the forward model in Eq.~\eqref{eq:near-field-intensity}.
The object parameters are updated using the initialization, data-fidelity term, and regularization described in Sec.~\ref{sec:reconstruction-procedure}.
The algorithm runs for 15000 iterations, with each angle visited approximately 10 times.
The maximum number of splats is set to 5 million.
The regularization weights and subvolume sizes are given in Appendix~\ref{app:optimization-settings}.

We compare BeST with an INR method for joint 3D reconstruction~\cite{gruenXRayNearFieldHolotomography2026a} and with a sequential workflow based on ASRM and filtered backprojection (FBP)~\cite{doraArtifactsuppressingReconstructionStrongly2024}.
The comparison methods use the parameter settings reported in the INR publication~\cite{gruenXRayNearFieldHolotomography2026a}.

\subsection{Far-Field Ptychotomography}
\label{sec:ptychography-methods}
The far-field ptychographic-tomography experiment was performed on a commercial STM32H753VIT6 microcontroller (STMicroelectronics), fabricated using \qty{40}{\nano\meter} technology~\cite{stmicroelectronicsSTM32H7Overview}.
The device contains a multilayer circuit semiconductor architecture comprising silicon, metallic interconnects, dielectric layers, and other fabrication materials, providing a structurally complex test specimen with features spanning multiple length scales.
A region of the exposed integrated-circuit die was selected for measurement.
The chip was thinned from the back before measurement.

	The experiment was conducted using the ptychographic nano-analytical microscope (PtyNAMi) at beamline P06 of PETRA~III at DESY in Hamburg, Germany~\cite{schroppPtyNAMiPtychographicNanoanalytical2020}.
	A schematic of the experimental configuration is shown on the right-hand side of Fig.~\ref{fig:experimental-setup}.
		At a photon energy of \qty{12.4}{\keV}, 30 diamond lenses~\cite{wangDiamondXrayLens2025} focused the incident beam to a focal spot with a full width at half maximum (FWHM) of approximately \qty{95}{\nano\meter} $\times$ \qty{135.8}{\nano\meter}.
The sample was positioned \qty{1.5}{\milli\meter} downstream of the focus, where the probe had an FWHM of \qty{847.9}{\nano\meter} $\times$ \qty{763.6}{\nano\meter}.
	An Eiger 4M detector with a pixel size of \qty{75}{\micro\meter} was positioned \qty{3.41}{\meter} downstream of the sample.
	This measurement geometry resulted in an isotropic reconstruction voxel size of \qty{8.8}{\nano\meter}.

The limited-angle tomographic dataset comprised 240 equally spaced projection angles spanning an angular range of \ang{120}.
At each angle, the specimen was scanned through the focused X-ray beam on a jittered $46\times46$ point grid, yielding 2116 diffraction patterns.

For far-field ptychography, we integrate the beta splats into CDTools~\cite{levitanCDTools}, a library that contains differentiable forward operators for two-dimensional ptychographic phase retrieval.
Using CDTools as the backend allows us to use the full feature set of state-of-the-art ptychographic forward models.
We replace the conventional two-dimensional object parameterization with the angle-dependent projections predicted by BeST.
The implementation uses $M=5$ probe modes and jointly optimizes the 3D object, probe modes, detector background, and scan positions.
The probe-mode model and detailed optimization settings are given in Appendices~\ref{app:probe-modes} and~\ref{app:optimization-settings}, respectively.

Scan-position refinement first estimates a global offset and then refines individual positions using Adam.
The number of splats is capped at 160 million; Appendix~\ref{app:optimization-settings} gives the refinement schedule and regularization settings.

For comparison with BeST, we applied a sequential workflow in which each projection was reconstructed independently using CDTools.
The projections are then initially aligned using scale-invariant feature transform alignment~\cite{Lowe2004} and refined using the procedure described in~\cite{Odstrcil2019}.
To assess differences in the recovered scan geometry, we compared the optimized probe positions from BeST with those from the independent projection reconstructions.
After per-angle similarity registration, we summarize the residual position discrepancies using their root-mean-square (RMS), median, and 95th-percentile magnitudes (Appendix~\ref{app:position-comparison}).
These residuals quantify differences that cannot be corrected by projection alignment restricted to translations as it is commonly done in tomographic reconstructions.

\section{Results}
\subsection{Near-Field Holography}
\begin{figure}
	\begin{subfigure}[t]{0.3\linewidth}
		\includegraphics[width=\linewidth]{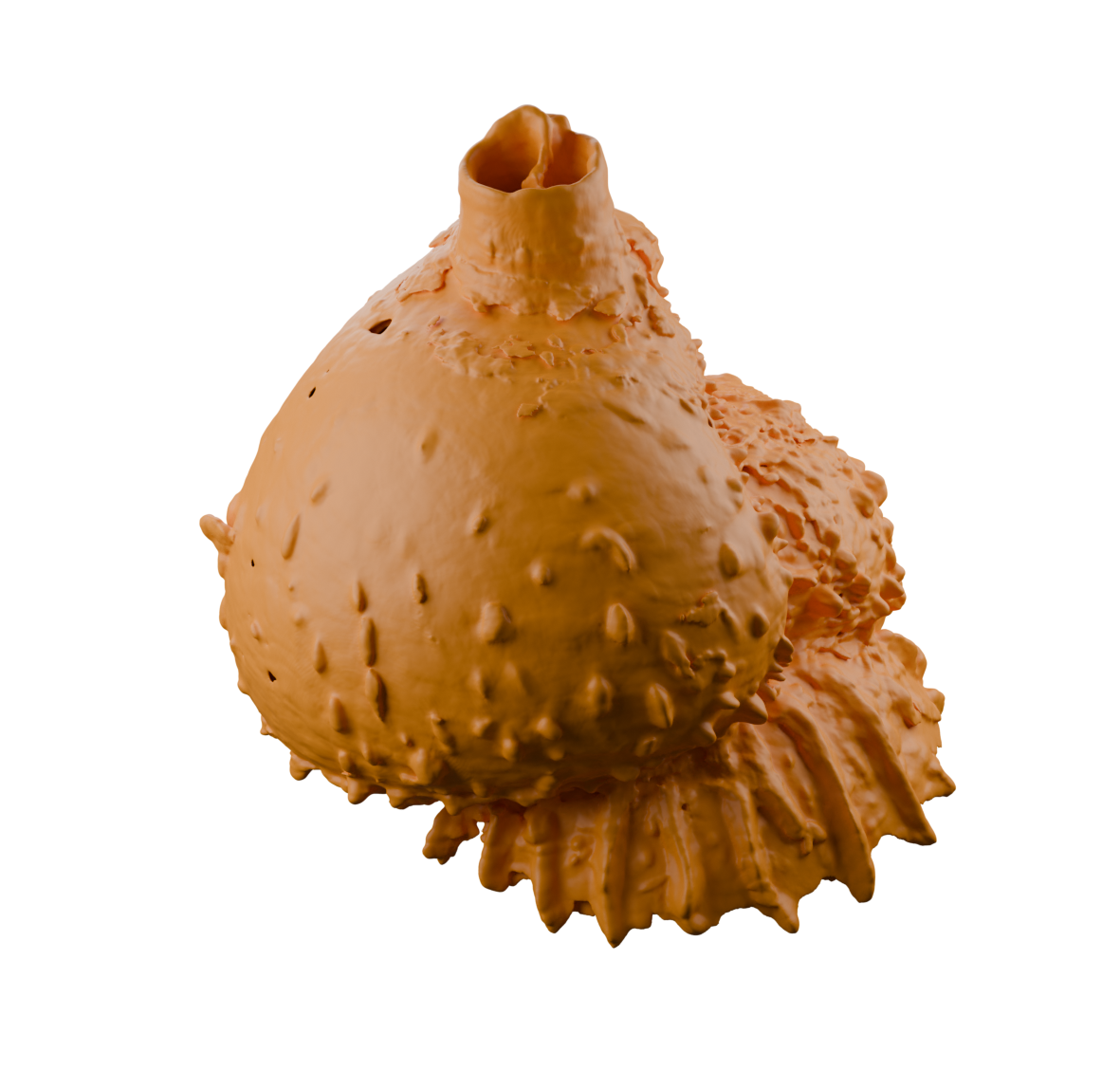}
	\end{subfigure}
	\begin{subfigure}[t]{0.695\linewidth}
		\includegraphics[width=\linewidth]{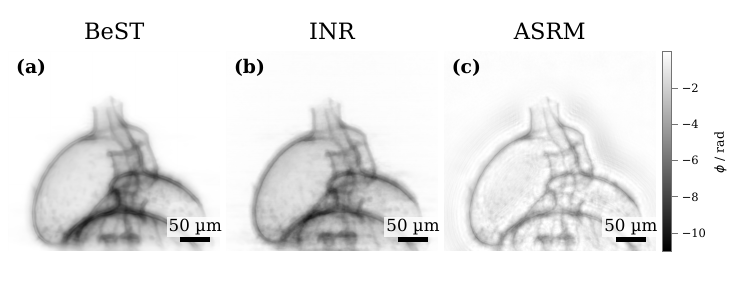}
	\end{subfigure}
	\caption{Left: 3D surface rendering of the foraminiferal specimen reconstructed with BeST.
	Right: Phase projections obtained with (a) BeST, (b) INR, and (c) ASRM, displayed on a common phase scale.
	The two joint methods show similar phase contrast, whereas the ASRM projection exhibits weaker broad-scale contrast and more pronounced fringes around the specimen.}
	\label{fig:near-field-projection}

\end{figure}
\begin{figure}
	\includegraphics[width=\linewidth]{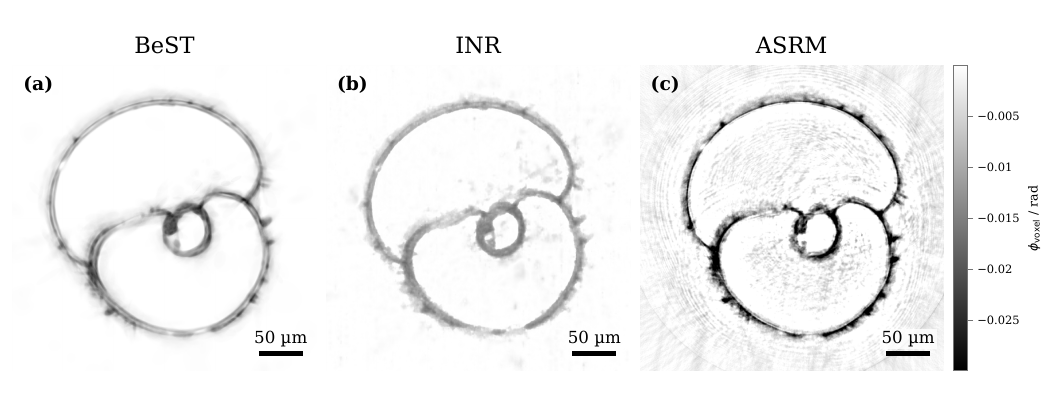}
	\caption{Corresponding transverse slices of the reconstructed phase shift for (a) BeST, (b) INR, and (c) the sequential ASRM-based reconstruction, displayed on a common phase scale.
	Two closely spaced bands are visible along portions of the wall in (a), particularly along the upper arc; their separation is less apparent in (b).
	The ASRM-based reconstruction shows pronounced concentric and streak-like artifacts within and around the specimen.}
	\label{fig:near-field-slice}
\end{figure}

Both joint methods (BeST and INR) recover similar low-frequency content, whereas the independently reconstructed ASRM projection exhibits less low-frequency content and more pronounced fringes around the specimen (Fig.~\ref{fig:near-field-projection}).
This observation is consistent with the additional constraints that a 3D object imposes on low-frequency components.

The corresponding transverse slices in Fig.~\ref{fig:near-field-slice} reveal differences in wall structure and background artifacts.
The BeST reconstruction shows comparatively uniform interiors and surrounding background, while the INR reconstruction shows broader wall profiles.
The sequential ASRM-based reconstruction exhibits strong wall contrast together with pronounced concentric and streak-like artifacts within and around the specimen.

Along portions of the wall, particularly the upper arc of the largest chamber, the BeST reconstruction separates two closely spaced bands that are less clearly distinguished in the INR reconstruction.
Ni et al.~\cite{niEarlyDiagenesisForaminiferal2020} reported calcite walls bordered by denser authigenic coatings on their inner and outer surfaces in diagenetically altered foraminifera.
This morphology provides a possible interpretation of the paired bands, although the reconstructed contrast alone does not establish their composition.

Reconstruction of the full $2048 \times 2048 \times 2048$ volume required approximately 30 minutes on a single NVIDIA RTX 6000 PRO, compared with approximately 24 hours for the INR approach and 28 hours for ASRM.
Distributing the work across multiple GPUs could enable reconstruction during acquisition and support experimental decisions based on the reconstructed structure.
The BeST reconstruction achieves a compression factor of 270 compared to a voxel representation, i.e. a reduction from 64 GiB to 242 MiB.

\subsection{Far-Field Ptychography}
\begin{figure}
	\includegraphics[width=\linewidth]{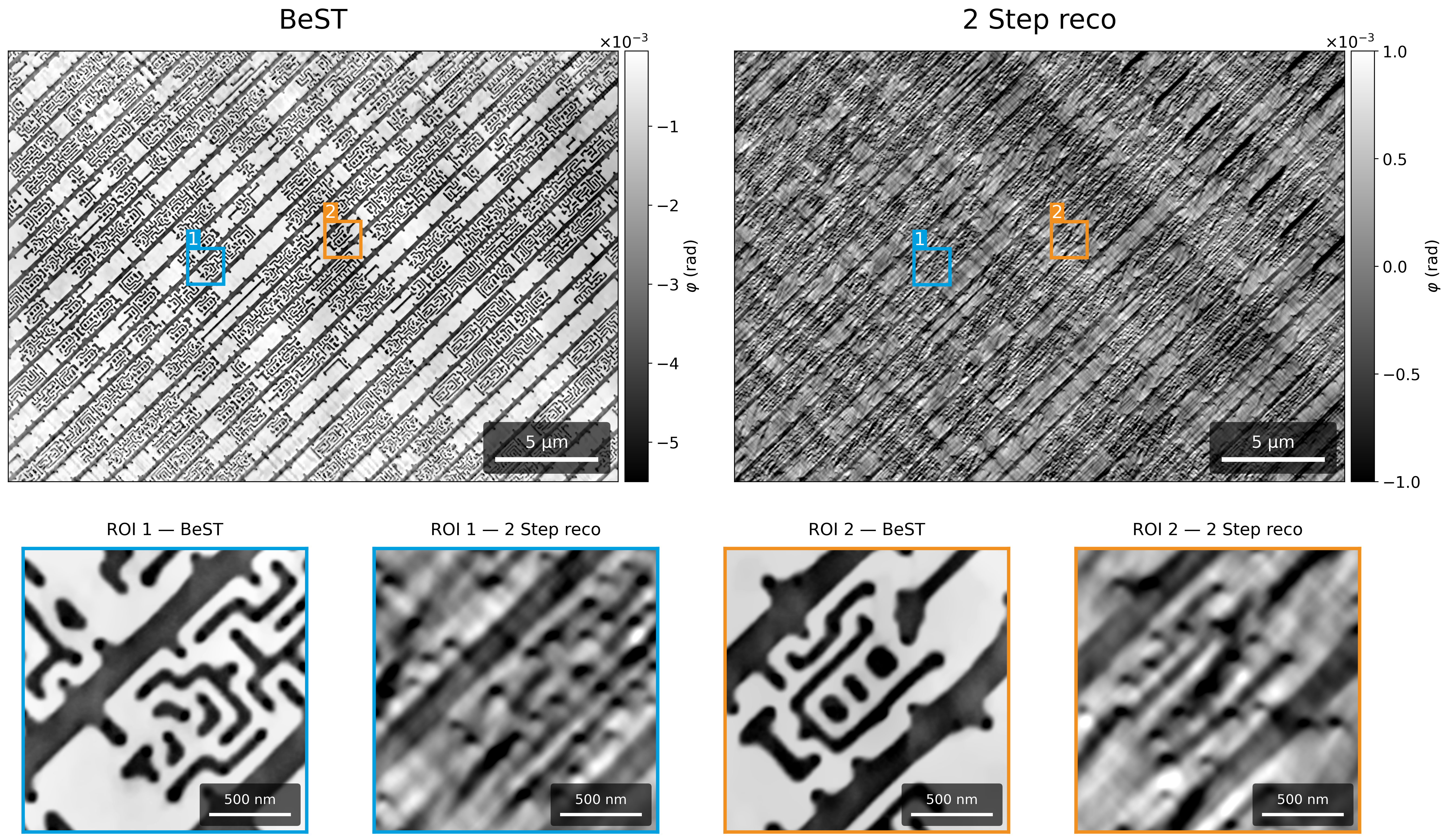}
	\caption{Reconstruction of metal layer~1 of the microchip.
	Top left: Slice obtained with BeST.
	Top right: Two-step reconstruction.
	Bottom: Enlarged views of ROI~1 and~2, with borders matching the corresponding boxes.
	}
	\label{fig:results-m1}
\end{figure}

Compared to holography, ptychography is a scanning technique and depends critically on accurately knowing the scan positions.
Ptychography optimizes the scan positions in each 2D projection separately, while a joint reconstruction stabilizes the scan position optimization by the common object.
This allows BeST to reconstruct repeated interconnect patterns (Fig.~\ref{fig:results-m1} top left), where the sequential reconstruction exhibits pronounced streaking and disruptions (Fig.~\ref{fig:results-m1} top right).
A further slice is shown in Fig.~\ref{fig:results-substrat} in Appendix~\ref{app:additional-ptychography} and a visualization showing a 3D rendering of the central region is in Visualization 1.

Local edge-sharpness yields an estimated FWHM of approximately \qty{23}{nm} for the BeST method (Appendix~\ref{app:edge-sharpness}).

To investigate the geometric differences between the workflows, we compared their recovered scan positions as described in Sec.~\ref{sec:ptychography-methods}.
The average RMS discrepancy of approximately 12.5~nm is consistent with scan-position differences contributing to projection inconsistencies and the resulting tomographic artifacts.
A more detailed analysis can be found in Fig.~\ref{fig:error} in Appendix~\ref{app:position-comparison}.

Reconstruction of the full $4800 \times 4800 \times 700$ voxel volume required 35~hours on a single NVIDIA H200 GPU.
The BeST reconstruction achieves a compression factor of 15 compared to a voxel representation, i.e. a reduction from 120 GiB to 7.7 GiB.

\section{Discussion \& Outlook}
The near-field results show that BeST can retain the benefits of a joint optimization while reducing reconstruction time relative to the INR approach.
This computational efficiency could support experimental decisions based on direct 3D reconstructions during acquisition.
For joint ptychography, BeST recovered interconnect structures in a dataset for which the two-step workflow produced reconstructions unsuitable for further analysis.

We introduce BeST, a method for joint phase retrieval and tomography based on compactly supported beta primitives with analytical projections.
The same continuous object representation supports both near-field holography and far-field ptychography and could be extended to other imaging modalities and tomography methods.
In the near-field holographic experiment BeST demonstrates faster reconstruction than the INR approach.
In the ptychographic experiment, joint refinement of scan positions under the constraint of a common 3D object reduces inconsistencies between projections, yielding fewer artifacts than the sequential reconstruction workflow.
Together, these results highlight beta splatting as a compact representation for direct 3D reconstruction, with benefits in computational efficiency, data storage, and reconstruction under measurement uncertainties.

\begin{backmatter}
\bmsection{Funding}
This work was supported by the German Federal Ministry of Research, Technology and Space (BMFTR) within the ErUM Data project CmarT (grant no.~13K2500189).
A. K. is supported by ESKAS Swiss Government Excellence Scholarship (2024.0479) and by the Ministry of Trade, Industry \& Energy Technology Innovation Program RS-2024-00419426, South Korea.
Part of this work was funded by Helmholtz Imaging, a platform of the Helmholtz Information \& Data Science Incubator, in the context of a visiting scientist position of M.G.-S.

\bmsection{Acknowledgments}
We thank Sha Ni for providing the foraminifera specimen.
We acknowledge support from Hi-Acts (Helmholtz Innovation Platform for Accelerator-based Technologies and Solutions).
We acknowledge the Maxwell computational resources operated at Deutsches Elektronen-Synchrotron DESY, Hamburg, Germany, which supported this research.
Parts of this research were carried out at beamlines P06 and P05.
Beamtime was allocated for the proposal ID I-20250626 and ID I-20230007.

\bmsection{Disclosures}
The authors declare no conflicts of interest.

\bmsection{Data and code availability}
The code and data supporting the results of this study will be made publicly available.
\end{backmatter}

\clearpage
\appendix
\setcounter{figure}{0}
\setcounter{equation}{0}
\renewcommand{\thefigure}{A\arabic{figure}}
\renewcommand{\theequation}{A\arabic{equation}}
\renewcommand{\theHfigure}{appendix.\arabic{figure}}
\renewcommand{\theHequation}{appendix.\arabic{equation}}
\section{Supporting Methods}

\subsection{Fresnel Scaling for Diverging Illumination}
\label{app:fresnel-scaling}
The Fresnel scaling theorem expresses propagation in the diverging illumination as an equivalent parallel-beam problem in coordinates referred to the object plane~\cite{paganinCoherentXrayOptics2006a}.
For the distances in Fig.~\ref{fig:experimental-setup}, the geometric magnification is $M_{\mathrm{geo}}=(z_{01}+z_{12})/z_{01}$, the transverse coordinate is $\mathbf{x}=\mathbf{x}_{\mathrm{det}}/M_{\mathrm{geo}}$, and the effective propagation distance is $z_{\mathrm{eff}}=z_{12}/M_{\mathrm{geo}}$.
The corresponding object-plane pixel size is $\Delta x=\Delta x_{\mathrm{det}}/M_{\mathrm{geo}}$, giving $\mathrm{Fr}_{\mathrm{eff}}=\Delta x^2/(\lambda z_{\mathrm{eff}})$.

\subsection{Multimode Ptychographic Forward Model}
\label{app:probe-modes}
To describe partial coherence and illumination instabilities, we represent the probe at each projection angle by $M=5$ mutually incoherent modes~\cite{thibaultReconstructingStateMixtures2013}.
For mode $m$, the exit wave at scan position $\mathbf{t}_{j,\varphi}$ is
\begin{align}
    \Psi^{\mathrm{exit}}_{m,j,\varphi}(\mathbf{x})
    = T_{\varphi}(\mathbf{x})
    P_{m,\varphi}(\mathbf{x}-\mathbf{t}_{j,\varphi}).
    \label{eq:appendix-mode-exit}
\end{align}
The predicted detector intensity is the incoherent sum of the mode intensities:
\begin{align}
    \mathcal{I}^{\mathrm{FF}}_{j,\varphi}(\boldsymbol{\nu})
    = a_{j,\varphi}^2\sum_{m=1}^{M}
    \left|\mathcal{F}\left[\Psi^{\mathrm{exit}}_{m,j,\varphi}\right](\boldsymbol{\nu})\right|^2
    +\mathcal{I}_{\mathrm{bg},\varphi}(\boldsymbol{\nu}).
    \label{eq:appendix-multimode-intensity}
\end{align}
The relative mode powers are contained in the norms of the complex probe fields and do not require separate normalized weights.
All modes interact with the same object transmission and share the scan positions at a given angle.

\subsection{Optimization Settings}
\label{app:optimization-settings}
\paragraph{Primitive refinement and densification.}
After 500 warm-up iterations, densification proceeds from iteration 500 to 15000.
Primitives with an accumulated average position-gradient magnitude of at least $10^{-5}$ are split if their largest scale exceeds $0.05\%$ of the largest reconstruction-volume dimension, and cloned otherwise.
Splitting replaces a parent with two children, whose positions are independently sampled from the parent's oriented Gaussian scale distribution.
Each child inherits the parent's rotation and shape parameter, with scales divided by $1.6$ and density coefficients halved.
Primitives are removed if either density coefficient falls below $10^{-6}$ or their centers lie outside the reconstruction bounds.
The initial learning rates for primitive position, scale, refractive and absorptive coefficients, and shape parameter are 0.0002, 0.001, 0.1, and 0.001, respectively, and are reduced linearly to 0.01 times their initial values during training.

\paragraph{Near-field regularization.}
The TV weight is set to $\lambda_{\mathrm{TV}} = 0.3$, and the gradient sampling cube $V_{\mathrm{TV}}$ has dimensions $128 \times 128 \times 128$.
On a separate cube, we apply $\ell_1$ regularization with weight $\lambda_{\ell_1} = 0.1$ to suppress the background.

\paragraph{Ptychographic optimization and scan-position refinement.}
The initial learning rate for the CDTools parameters is 0.1 and is reduced to 0.01 after 20000 iterations.
In our implementation, scan-position refinement proceeds in two stages.
Between iterations 500 and 2000, a global offset is evaluated using 40 randomly selected central scan positions, with candidate offsets in the range $(-200,200)$ and an acceptance threshold of at least $5\%$ improvement in the loss.
Subsequently, individual scan positions are refined using Adam.
The regularization weights are set to $\lambda_{\ell_1} = 0$ and $\lambda_{\mathrm{TV}} = 1$, and the queried subcube has dimensions $256 \times 256 \times 256$.

\subsection{Comparison of Recovered Scan Positions}
\label{app:position-comparison}
For each projection angle, the corresponding point clouds were registered using the Umeyama algorithm~\cite{umeyamaLeastSquares1991}, allowing translation, rotation, and uniform scaling.
The remaining position discrepancies were summarized by their root-mean-square (RMS) magnitude, median magnitude, and 95th percentile at each angle.
These residuals quantify differences not explained by a global similarity transformation and therefore cannot be corrected by projection alignment restricted to translations.
Let $\mathbf{u}_{j,\varphi}$ and $\mathbf{v}_{j,\varphi}$ denote the positions recovered by the independent projection reconstructions and BeST, respectively.
After fitting a scale $c_{\varphi}$, rotation $\mathbf{Q}_{\varphi}$, and translation $\mathbf{d}_{\varphi}$, the residual magnitude is
\begin{align}
    e_{j,\varphi}
    =\left\|c_{\varphi}\mathbf{Q}_{\varphi}\mathbf{u}_{j,\varphi}
    +\mathbf{d}_{\varphi}-\mathbf{v}_{j,\varphi}\right\|_2.
\end{align}
For $N_{\varphi}$ scan positions, the RMS discrepancy is
$\sqrt{N_{\varphi}^{-1}\sum_j e_{j,\varphi}^2}$;
the median and 95th percentile are calculated from the same residual magnitudes.

\begin{figure}[htbp]
    \centering
    \includegraphics[width=0.75\linewidth]{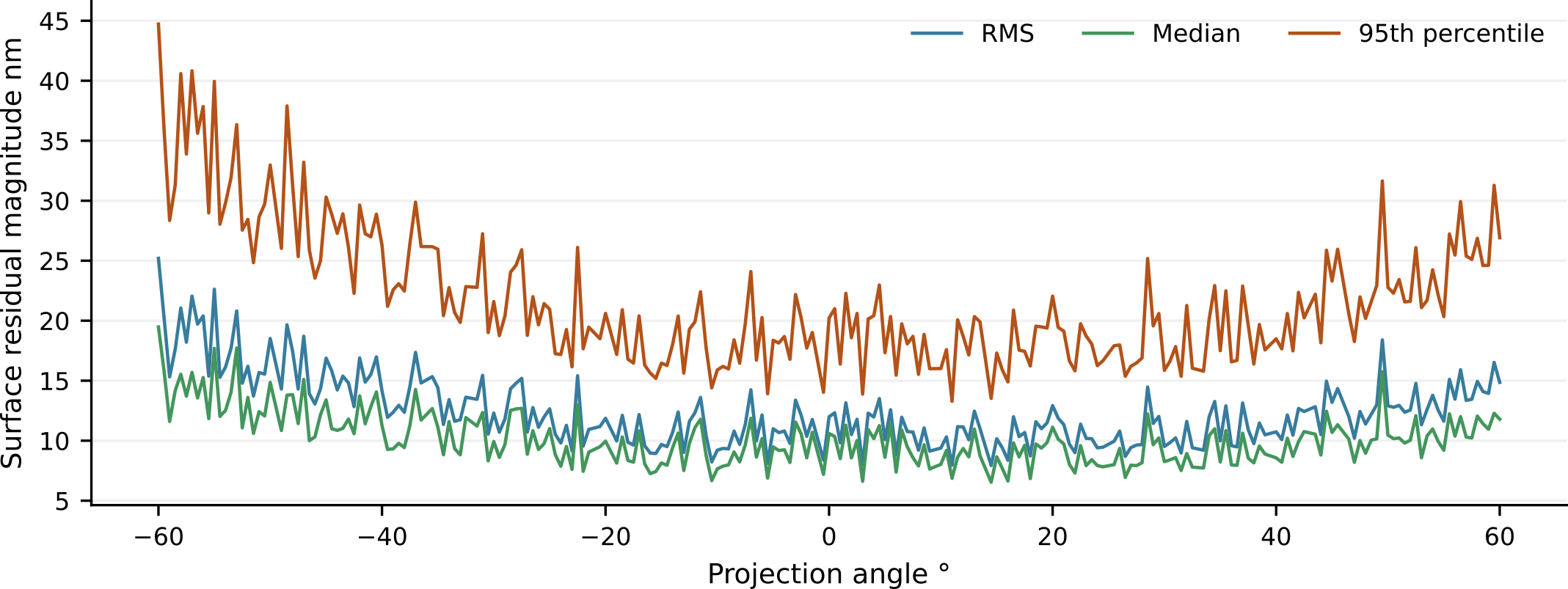}
    \caption{Scan-position discrepancies between BeST and the independent projection reconstructions after per-angle similarity registration.
    RMS, median, and 95th-percentile residual position magnitudes are shown as a function of projection angle, calculated as described in this appendix.
    The residuals indicate geometric differences between the two reconstructions that are not explained by translation, rotation, and uniform scaling.}
    \label{fig:error}
\end{figure}

\subsection{Evaluation of the Inverse Abel Transform}
\label{app:abel-derivation}
For a fixed shape parameter, write $p=p(b)$.
Differentiating the projected kernel within its support gives
$\partial_r\mathcal{B}_2(r;b)=-2pr(1-r^2)^{p-1}$.
Substitution into Eq.~\eqref{eq:inverse-abel} yields
\begin{align}
    \mathcal{B}_3(R;b)
    =\frac{2p}{\pi}\int_R^1
    \frac{r(1-r^2)^{p-1}}{\sqrt{r^2-R^2}}\,\mathrm{d}r.
\end{align}
For $0\leq R<1$, the substitution $u=(r^2-R^2)/(1-R^2)$ gives
\begin{align}
    \mathcal{B}_3(R;b)
    &=\frac{p}{\pi}(1-R^2)^{p-\frac12}
    \int_0^1 u^{-\frac12}(1-u)^{p-1}\,\mathrm{d}u\\
    &=\frac{p}{\pi}(1-R^2)^{p-\frac12}
    \frac{\Gamma(\tfrac12)\Gamma(p)}{\Gamma(p+\tfrac12)}.
\end{align}
The integral is Euler's beta integral.
Using $\Gamma(\tfrac12)=\sqrt{\pi}$ and $p\Gamma(p)=\Gamma(p+1)$ gives Eq.~\eqref{eq:beta-kernel-3d}.
The boundary value follows by continuity because $p>\tfrac12$.

\clearpage
\subsection{Slice below Metal-Layer 1}
\label{app:additional-ptychography}
\begin{figure}[h!]
	\includegraphics[width=\linewidth]{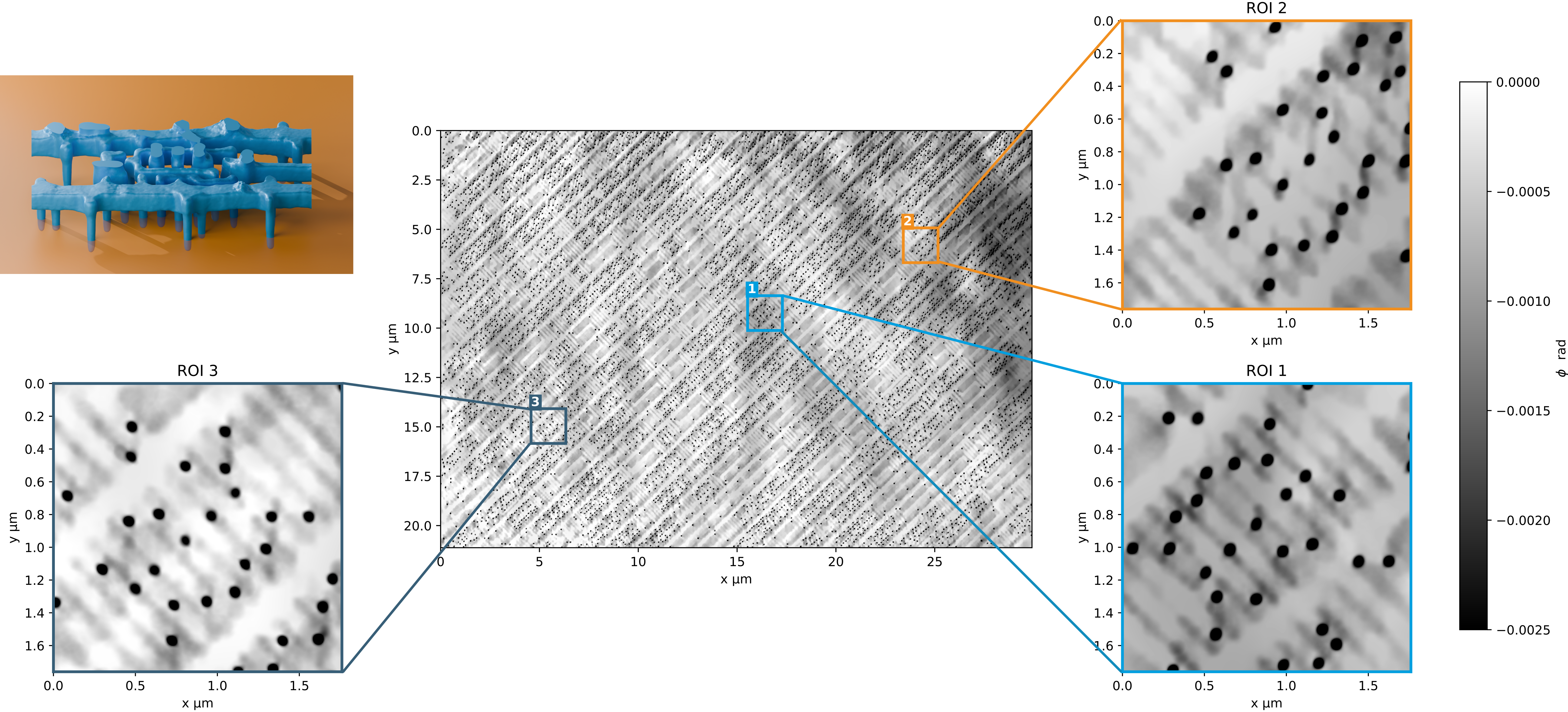}
	\caption{Reconstruction of a slice below metal-layer 1 of the microchip.
	The central panel shows the slice overview, and ROIs~1--3 show enlarged regions at the corresponding lateral positions indicated in Fig.~\ref{fig:results-m1}.
	Distinct contact plugs appear as dots against a spatially varying background.}
	\label{fig:results-substrat}
\end{figure}

\subsection{Local Edge-Sharpness Estimates}
\label{app:edge-sharpness}
\begin{figure}[htbp]
    \centering

    \begin{subfigure}[t]{0.49\textwidth}
        \centering
        \includegraphics[width=\linewidth]{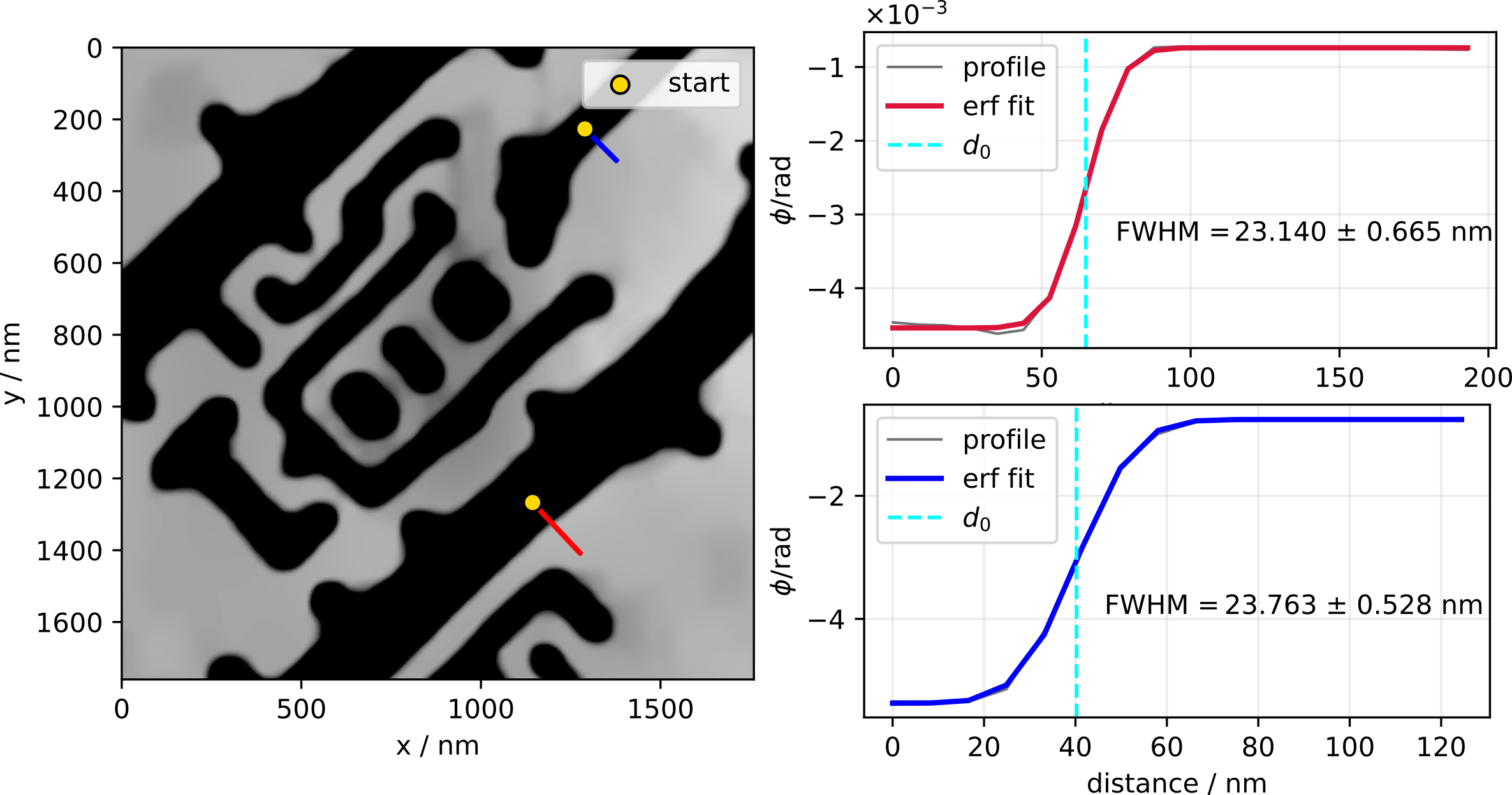}
        \caption{Selected edges in the metal-layer slice shown in Fig.~\ref{fig:results-m1}.}
        \label{fig:fwhm-upper}
    \end{subfigure}
    \begin{subfigure}[t]{0.49\textwidth}
        \centering
        \includegraphics[width=\linewidth]{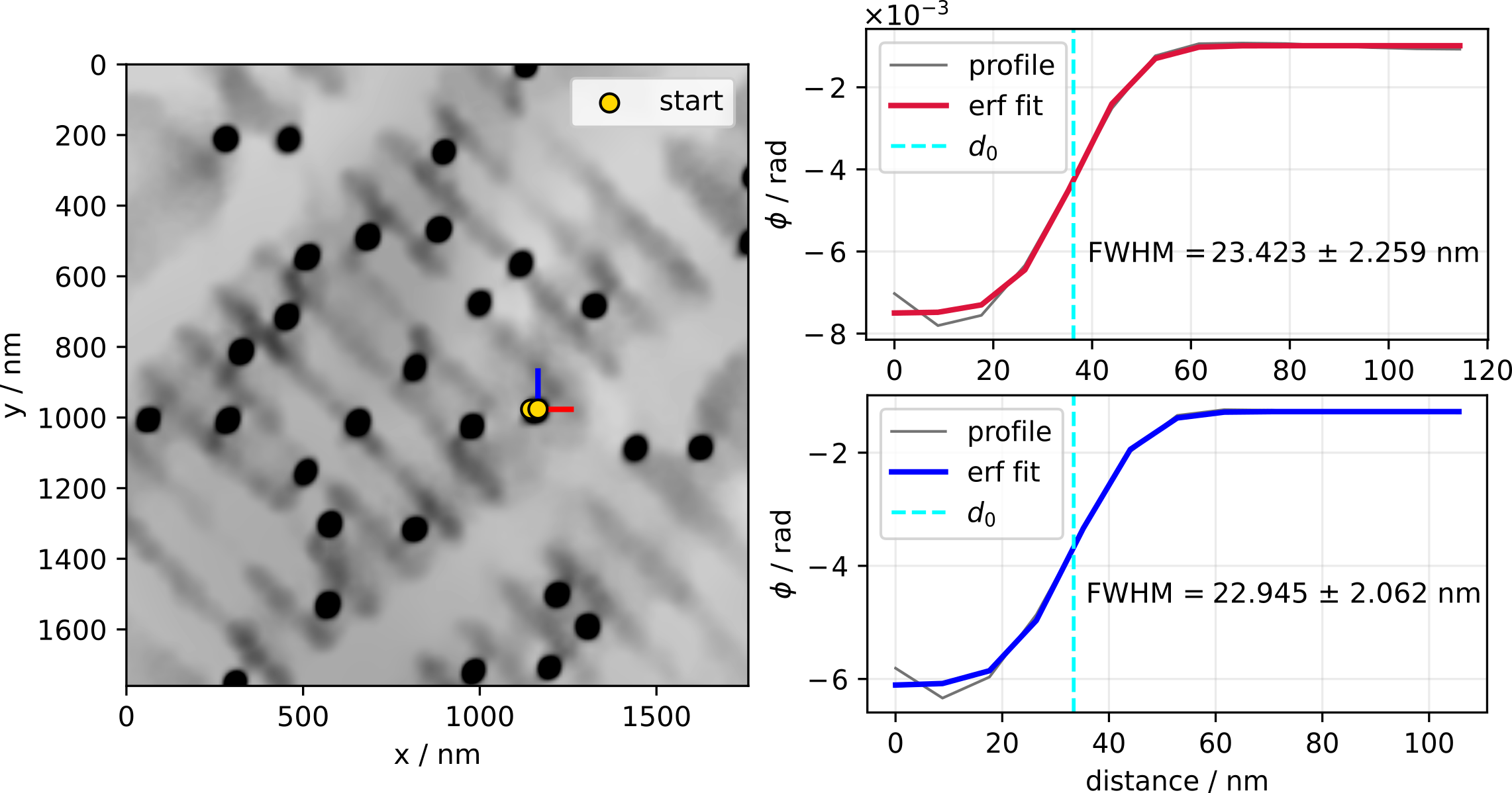}
        \caption{Selected edges in the slice below the metal-layer 1 shown in Fig.~\ref{fig:results-substrat}.}
        \label{fig:fwhm-lower}
    \end{subfigure}

    \caption{
        Local edge-sharpness analysis of the ptychographic reconstruction obtained with BeST.
        In each panel, the left image marks the sampled edges with red and blue segments and yellow starting points.
        The plots on the right show the phase profiles (gray), error-function fits (matching colors), and fitted edge centers $d_0$ (cyan dashed lines).
        The reported FWHM is that of the Gaussian line-spread function corresponding to each fitted edge response.
    }
    \label{fig:fwhm}
\end{figure}

Figure~\ref{fig:fwhm} assesses local edge sharpness in metal layer 1 and in a slice below it using error-function fits to selected phase profiles.
The FWHM values of the corresponding Gaussian line-spread functions range from \qty{23.1}{\nano\meter} to \qty{23.8}{\nano\meter} in the metal-layer slice and from \qty{22.9}{\nano\meter} to \qty{23.4}{\nano\meter} in the slice below.
These measurements characterize selected local edges and do not establish uniform resolution throughout the volume.

\clearpage
\bibliography{./bib}

\end{document}